\documentclass[english,pra,showpacs,reprint,superscriptaddress]{revtex4-1}
\usepackage[latin9]{inputenc}
\usepackage{color}
\usepackage{babel}
\usepackage{verbatim}
\usepackage{bm}
\usepackage{amsmath}
\usepackage{amssymb}
\usepackage{graphicx}
\usepackage{tikz}
\usepackage[unicode=true,pdfusetitle,
 bookmarks=true,bookmarksnumbered=false,bookmarksopen=false,
 breaklinks=false,pdfborder={0 0 0},pdfborderstyle={},backref=false,colorlinks=true]
 {hyperref}
\hypersetup{citecolor=blue}

\makeatletter

\usepackage{epstopdf}
\usepackage{amsfonts}
\usepackage{mathrsfs}\usepackage{bm}\usepackage{appendix}\usepackage{color}
\usepackage{txfonts}
\usepackage{float}
\usepackage{epstopdf}\usepackage{ulem}

\makeatother

\begin{document}

\title{Transport tuning of photonic topological edge states by optical cavities}
\author{Chang-Yin Ji}
\altaffiliation{These two authors contribute equally to this work.}
\affiliation{Key Lab of advanced optoelectronic quantum architecture and measurement (MOE), Beijing Key Lab of Nanophotonics $\&$ Ultrafine Optoelectronic Systems,
and School of Physics, Beijing Institute of Technology, Beijing 100081,
China}
\affiliation{China Academy of Engineering physics, Mianyang, Sichuan, 621900, China}
\author{Gui-Bin Liu}
\altaffiliation{These two authors contribute equally to this work.}
\affiliation{Key Lab of advanced optoelectronic quantum architecture and measurement (MOE), Beijing Key Lab of Nanophotonics $\&$ Ultrafine Optoelectronic Systems,
and School of Physics, Beijing Institute of Technology, Beijing 100081,
China}
\author{Yongyou Zhang}
\email[Corresponding author: ]{yyzhang@bit.edu.cn}
\affiliation{Key Lab of advanced optoelectronic quantum architecture and measurement (MOE), Beijing Key Lab of Nanophotonics $\&$ Ultrafine Optoelectronic Systems,
and School of Physics, Beijing Institute of Technology, Beijing 100081,
China}

\author{Bingsuo Zou}
\affiliation{Key Lab of advanced optoelectronic quantum architecture and measurement (MOE), Beijing Key Lab of Nanophotonics $\&$ Ultrafine Optoelectronic Systems,
and School of Physics, Beijing Institute of Technology, Beijing 100081,
China}
\author{Yugui Yao}
\email{ygyao@bit.edu.cn}
\affiliation{Key Lab of advanced optoelectronic quantum architecture and measurement (MOE), Beijing Key Lab of Nanophotonics $\&$ Ultrafine Optoelectronic Systems,
and School of Physics, Beijing Institute of Technology, Beijing 100081,
China}

\date{\today}
\begin{abstract}

Crystal-symmetry-protected photonic topological edge states (PTESs) based on air rods in conventional dielectric materials are designed as photonic topological waveguides (PTWs) coupled with side optical cavities. We demonstrate that the cavity coupled with the PTW can change the reflection-free transport of the PTESs, where the cavities with single mode and twofold degenerate modes are taken as examples. The single-mode cavities are able to perfectly reflect the PTESs at their resonant frequencies, forming a dip in the transmission spectra. The dip full width at half depth depends on the coupling strength between the cavity and PTW and thus on the cavity geometry and distance relative to the PTW. While the cavities with twofold degenerate modes lead to a more complex PTES transport whose transmission spectra can be in the Fano form. These effects well agree with the one-dimensional PTW-cavity transport theory we build, in which the coupling of the PTW with cavity is taken as $\delta$ or non-$\delta$ type. Such PTWs coupled with side cavities, combining
topological properties and convenient tunability, have wide diversities
for topological photonic devices.
\end{abstract}


\maketitle

\section{Introduction}

Discovery of electronic topological systems has revolutionized fundamental
cognition of phase transitions in condensed matter physics \cite{klitzing1980new, haldane1988model, kane2005quantum, bernevig2006quantum, fu2007topological, zhang2009topological, moore2010birth, hasan2010colloquium, wan2011topological, burkov2011topological, qi2011topological, chang2013experimental}.
The fascinating topological phases have been extended to the fields
of electromagnetic waves, in which the optical analogues of quantum
Hall (QH) and quantum spin Hall (QSH) effects can be observed \cite{haldane2008, wang2008reflection,yu2008one,wang2009observation, khanikaev2013photonic, chen2014experimental, ma2015guiding, cheng2016robust, he2016photonic, slobozhanyuk2017three,bahari2017nonreciprocal,wu2015scheme, xu2016accidental, anderson2017unidirectional, zhu2017topological, hafezi2011robust, hafezi2013imaging, lu2014topological, lu2016symmetry}.
The QH photonic topological insulators (PTIs) were theoretically
designed \cite{haldane2008, wang2008reflection} and soon afterwards
experimentally implemented \cite{wang2009observation}, which are
consisted of gyromagnetic materials with an applied magnetic field
to break the time reversal symmetry (TRS). Oppositely, for the QSH
PTIs, the key point is to achieve the Kramer's degeneracy by a kind
of pseudo-TRS. Different from the spin-$\frac{1}{2}$ electronic systems,
real TRS in photonic systems cannot ensure the Kramer's degeneracy,
for which additional symmetry is required. For example, pseudo-spin states can be implemented by utilizing clockwise
and anticlockwise modes in the coupling rings \cite{hafezi2011robust}, hybridization of transverse electric and magnetic waves \cite{chen2014experimental, khanikaev2013photonic, cheng2016robust, ma2015guiding, dong2017valley,
slobozhanyuk2017three, he2016photonic},
or degeneracy of Bloch modes due to crystal symmetry \cite{wu2015scheme, barik2016two,
xu2016accidental, anderson2017unidirectional, zhu2017topological, he2016acoustic, mei2016pseudo, zhang2017topological, xia2017topological, he2016acoustic,PhysRevLett.120.217401,brendel2018snowflake,gorlach2018far,yves2017crystalline,
barik2018topological}.  All these systems possess
topologically protected edge states which are robust against defects
to support reflection-free transport of photons.

As well known, waveguides are very important devices in photonics,
along which photons are transported to carry information. Cavities
coupled with waveguides are usually designed to control the transport
of light, forming traps, filters, and switches \cite{PhysRevLett.96.153601, villeneuve1996single, fan1998channel, fan2002sharp, wang2003compact, nozaki2010sub, dong2014multi, hu2018transmission, jiangsingle}.
The emergence of PTIs provides just right chances to realize reflection-free
photonic topological waveguides (PTWs) using photonic topological
edge states (PTESs). Previous studies about PTIs mainly focused on
how to achieve the topological photonic systems and to demonstrate the
robustness of the PTESs. What will happen if the PTWs are coupled with optical defects (for example, a cavity)? It is still a fascinating subject to uncover, though researchers have realized that optical defects can flip the photonic pseudo spin \cite{gao2016probing}. We in this work demonstrate that the generally believed reflection-free PTESs can be changed by coupled cavities as their eigenfrequencies lie within the PTI gap. A realistic aspect is revealed to turn on and off the transport of the PTESs. The transmission spectra have abundant line shapes
near the resonant frequency of the cavity, which agrees well with the one-dimensional PTW-cavity transport theory that we build. The bend immune PTESs, combining the cavities that can flip the pseudospin of the PTESs, provide wide diversities for photonic systems, such as topological optical switches, filters, and logic gates.

This work is organized as follows. In Sec.~II, we first design a crystal symmetry protected (CSP) PTI based on air hole lattice in a silicon substrate and then show the PTESs are free backscattering for bent edges. In Sec.~III, two types of optical cavities are used to tuning the transport of the PTESs. At last, a conclusion is summarized in Sec.~IV.

\section{Crystal symmetry protected PTIs}

\begin{figure*}
\centering
\includegraphics[width=1.0\textwidth]{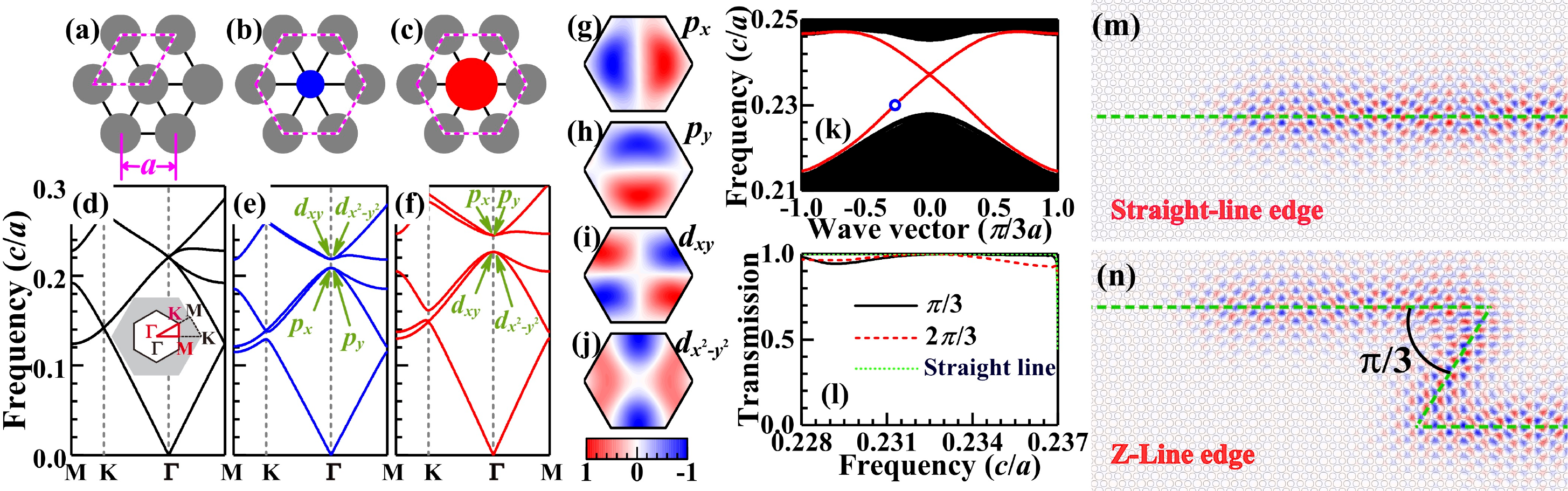}
\caption{Unit cell of air-rod triangle lattices with lattice
constant $\sqrt{3}a$ and air rod radius $r=0.35a$ (a), and those
with the radius of centric rod being shrunk to $0.25a$ (b) and expanded
to $0.45a$ (c). (d-f) Band structures of the photonic crystals, corresponding to
(a-c). The inset in (d) shows the BZs for the primitive and enlarged cells.
(g-j) Electric field distributions of
the modes $p_{x}$, $p_{y}$, $d_{xy}$, and $d_{x^{2}-y^{2}}$. (k)
Photonic band structure of the bulk states (black lines) and edge
states (red lines) between the PTI and trivial photonic insulator.
(l) Transmission spectra of the pseudospin-up edge states for interfaces
with different bending angles, of which electric field distributions
for straight-line and Z-line (bending angle is $\pi/3$) edges at the frequency marked by
the blue circle in (k) are shown in (m) and (n), respectively. Green
lines in (m, n) represent the interfaces between the PTI (below) and
trivial photonic insulator (above). For (k-n), the radii of the small
and large air rods are $0.35a$ and $0.45a$ for the PTI, while are $0.32a$
and $0.42a$ for the trivial photonic insulator, and they have the same lattice constant of $\sqrt{3}a$. These values guarantee
the match between the bulk band gaps for the PTI and trivial photonic crystal.}
\label{structures}
\end{figure*}

Recently, a scheme using dielectric materials was proposed
to achieve CSP PTIs \cite{wu2015scheme},
which have been realized on various platforms \cite{wu2015scheme, barik2016two,
xu2016accidental, anderson2017unidirectional, zhu2017topological, he2016acoustic, mei2016pseudo, zhang2017topological, xia2017topological, he2016acoustic,PhysRevLett.120.217401,brendel2018snowflake,gorlach2018far,yves2017crystalline,
barik2018topological}. For example, microwaves \cite{yves2017crystalline}, infrared range \cite{barik2016two,gorlach2018far, barik2018topological}, and visible range. We first design a CSP
PTI and then study how to tune the transport of its PTESs
with cavities, and only consider the transverse magnetic (TM) waves whose electric and magnetic
fields are out of and in $xy$-plane, respectively. The bands and transport properties of the TM waves are solved within the finite element method (FEM) by the code of COMSOL Multiphysics. Our designed CSP PTI originates from a primitive triangular lattice
of air rods on a common dielectric substrate. Without loss of generality, we take silicon as an example, with relative dielectric
constant $\varepsilon_{r}=11.7$ \cite{smith1985handbook}. The air-rod lattices are practical in experiments with advanced micro-/nano-fabrication technologies \cite{birner2001silicon}, showing advantage over the gyromagnetic and bianisotropy materials in optical frequency region. The primitive lattice constant, $a$, is the distance between the centers of two neighbor air rods and
the air rod radius is $r=0.35a$ [see Fig.~\ref{structures}(a)]. Then the primitive cell [pink dashed parallelogram] is enlarged to three times
large [hexagon in Fig.~\ref{structures}(b)], forming the present cells with lattice constant $\sqrt{3}a$. Accordingly, the Dirac cones
at the high symmetry points $\bm K$ and $\bm K'$ of the primitive Brillouin
zone (BZ) are folded to the $\bm \Gamma$ point of the present BZ, resulting in
fourfold degenerate states at $\bm \Gamma$ point [see Fig.~\ref{structures}(d) and its inset].
This fourfold degeneracy can be broken by decreasing or increasing
the radius of the centric air rod [see Figs.~\ref{structures}(b-c)]. Because all the structures in Figs.~\ref{structures}(a-c)
keep the $C_{6v}$ symmetry which has two 2D irreducible
representations of $E_{1}$ and $E_{2}$, the fourfold degeneracy splits
into two twofold degeneracies, as shown in Figs.~\ref{structures}(e-f). Analogous to electronic systems, two
bases of $E_{1}\ (E_{2})$ are $p_{x}$ and $p_{y}$ ($d_{xy}\ {\rm and}\ d_{x^{2}-y^{2}}$)
orbitals, whose electric field distributions are in Figs.~\ref{structures}(g-j). Orbital projection results of the
bands are similar to those in Ref. \cite{wu2015scheme}. For the case
with decreasing centric rod [see Fig.~\ref{structures}(b)] the frequencies
of the $p_{x}$ and $p_{y}$ orbitals are lower than those of the
$d_{xy}$ and $d_{x^{2}-y^{2}}$ orbitals [see Fig.~\ref{structures}(e)]
without band inversion, being a trivial photonic insulator. While
for the case with increasing centric rod [see Fig.~\ref{structures}(c)]
the frequencies of the $p_{x}$ and $p_{y}$ orbitals are higher than
those of the $d_{xy}$ and $d_{x^{2}-y^{2}}$ orbitals [see Fig.~\ref{structures}(f)],
which implies a band inversion and hence a nontrivial PTI. The principle
of this nontrivial topology connects with that of $Z_{2}$
electronic topological insulators protected by TRS. Here, the 2D irreducible representations of $E_{1}$ and $E_{2}$ provide opportunities to construct
a pseudo-TRS and hence Kramer's doubly degenerate states. Recombination of the four orbitals
provides pseudospin states of the system \cite{wu2015scheme,xu2016accidental,anderson2017unidirectional,zhu2017topological,he2016acoustic,mei2016pseudo,zhang2017topological,xia2017topological,PhysRevLett.120.217401,brendel2018snowflake},
namely,
\begin{align}
p_{\pm}=(p_{x}\pm ip_{y})/\sqrt{2},\quad d_{\pm}=(d_{x^{2}-y^{2}}\pm id_{xy})/\sqrt{2},\label{pseudospinstates}
\end{align}
where $p_{+}$ ($d_{+}$) and $p_{-}$ ($d_{-}$) are the pseudospin-up
and -down states of the $p$ ($d$) band, respectively. According to
Ref.~\cite{wu2015scheme}, the pseudo-TRS operator, ${\cal T}$,
can be expressed as ${\cal T=UK}$ where ${\cal U}=-\sigma_{y}$ ($\sigma_{y}$
is the Pauli matrix operated on bases $p_{\pm}$ or $d_{\pm}$) and
${\cal K}$ is the complex conjugate operator. It is direct to check
${\cal T}{}^{2}=-1$ on the bases of $p_{\pm}$ or $d_{\pm}$. This
pseudo-TRS in the present photonic system guarantees the nontrivial
$Z_{2}$ topology of the structure in Fig.~\ref{structures}(c).

Figure \ref{structures}(k) shows the bands of the helical edge states localized
at the interface between the PTI and trivial photonic insulator as
the red lines. When excited by a pseudospin-polarized source, the
PTES propagates only in one direction and has negligible back-reflection along the bending PTW. Figures \ref{structures}(m) and \ref{structures}(n)
show the rightward moving electric field excited by a pseudospin-up
source at the frequency of 0.23$c/a$ [blue circle dot in Fig.~\ref{structures}(k)]
for two kinds of PTWs, one of which is straight and the other is
Z-type with $\pi/3$ bending angle. The transmission spectra calculated by the scattering matrix method \cite{lu2016valleytransport} [see Appendix A] are shown in Fig.~\ref{structures}(l) where another Z-type edges with bending
angles $2\pi/3$ is also given. All the transmissivities
approach 100\% within the band gap of the PTI, indicating
the nontrivial topology of the photonic crystals. There is a tiny gap at the cross point of the two-branch
edge bands [too tiny to be visible in Fig.~\ref{structures}(k)],
whose value can be tuned by changing the geometry of the edge interface \cite{wu2015scheme, xu2016accidental, anderson2017unidirectional, zhu2017topological, he2016acoustic, mei2016pseudo, zhang2017topological, xia2017topological, he2016acoustic, PhysRevLett.120.217401,brendel2018snowflake}, being about 0.00021$c/a$ in our structure. The tiny gap essentially originates from the breaking of the $C_{6v}$ symmetry at the interface. In certain cases, it can disappear if the mirror and chiral symmetries are both satisfied simultaneously \cite{kariyado2017t}. The $C_{6v}$ symmetry is responsible for the emergence of the pseudospin states in Eq.~(\ref{pseudospinstates}), as well as the pseudo-TRS $\mathcal{T}$, therefore any breaking of
the $C_{6v}$ symmetry may destroy the topological properties of the systems. Moreover, certain defects breaking the $C_{6v}$ symmetry would change the system's
topological properties, including the reflection-free transport of
the PTES.

\section{Tuning PTES Transport}

\begin{figure}
\centering
\includegraphics[width=0.4\textwidth]{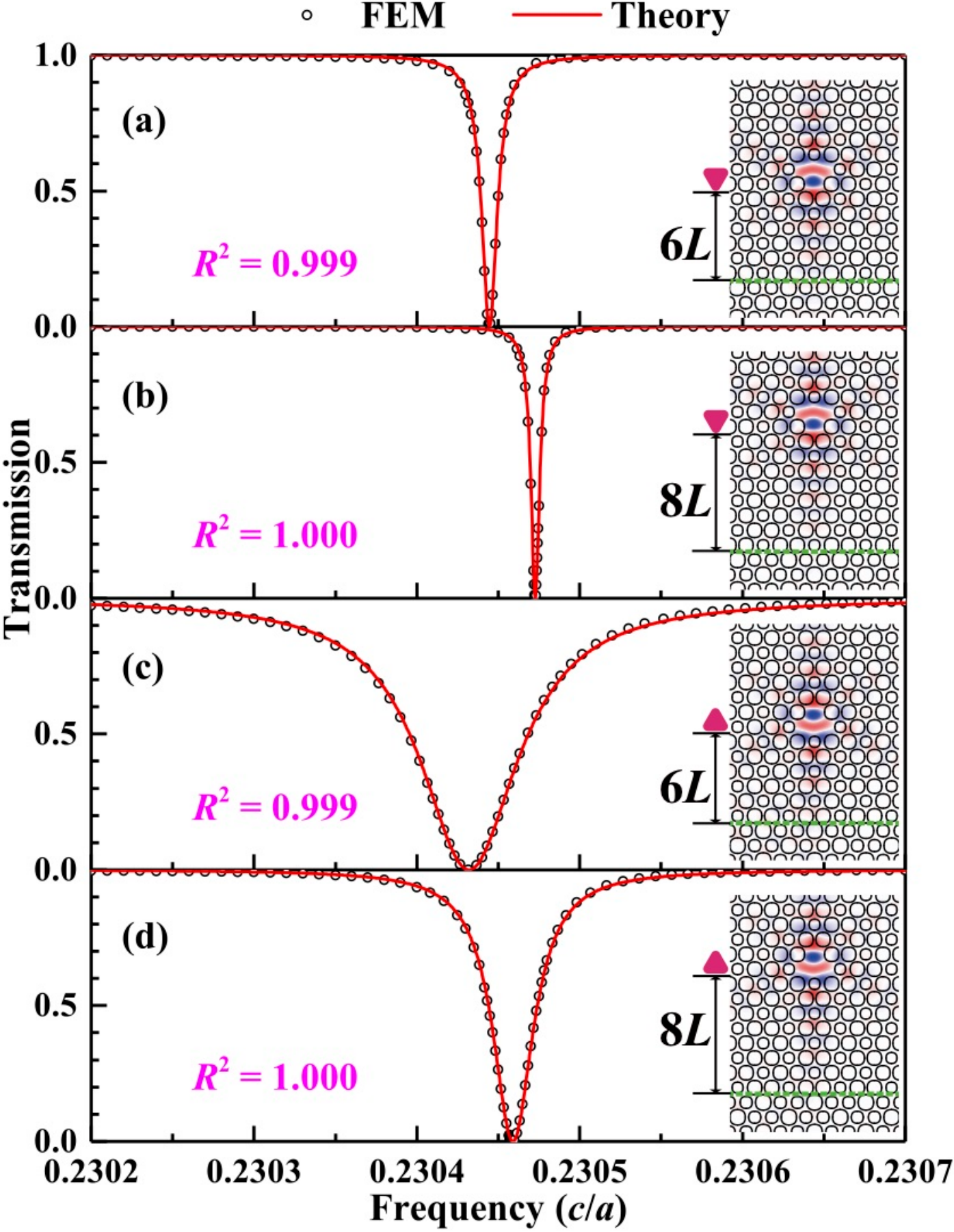}
\caption{Transmission spectra of the topological edges that are coupled with
side single-mode cavities for four different cases: (a, b) $\bigtriangledown$-type
cavities and (c, d) $\bigtriangleup$-type cavities. All cavities are achieved by deleting two bigger and one smaller air rods. Notations
of $6L$ and $8L$ show the distances between the cavities and edges,
i.e., there are 6 and 8 layers, respectively. Circle dots are calculated
in FEM by the COMSOL code, while red lines are the fitted curves whose
fitting precisions are described by the adjusted $R$-square.}
\label{cavity1}
\end{figure}

\begin{figure}
\centering
\includegraphics[width=0.4\textwidth]{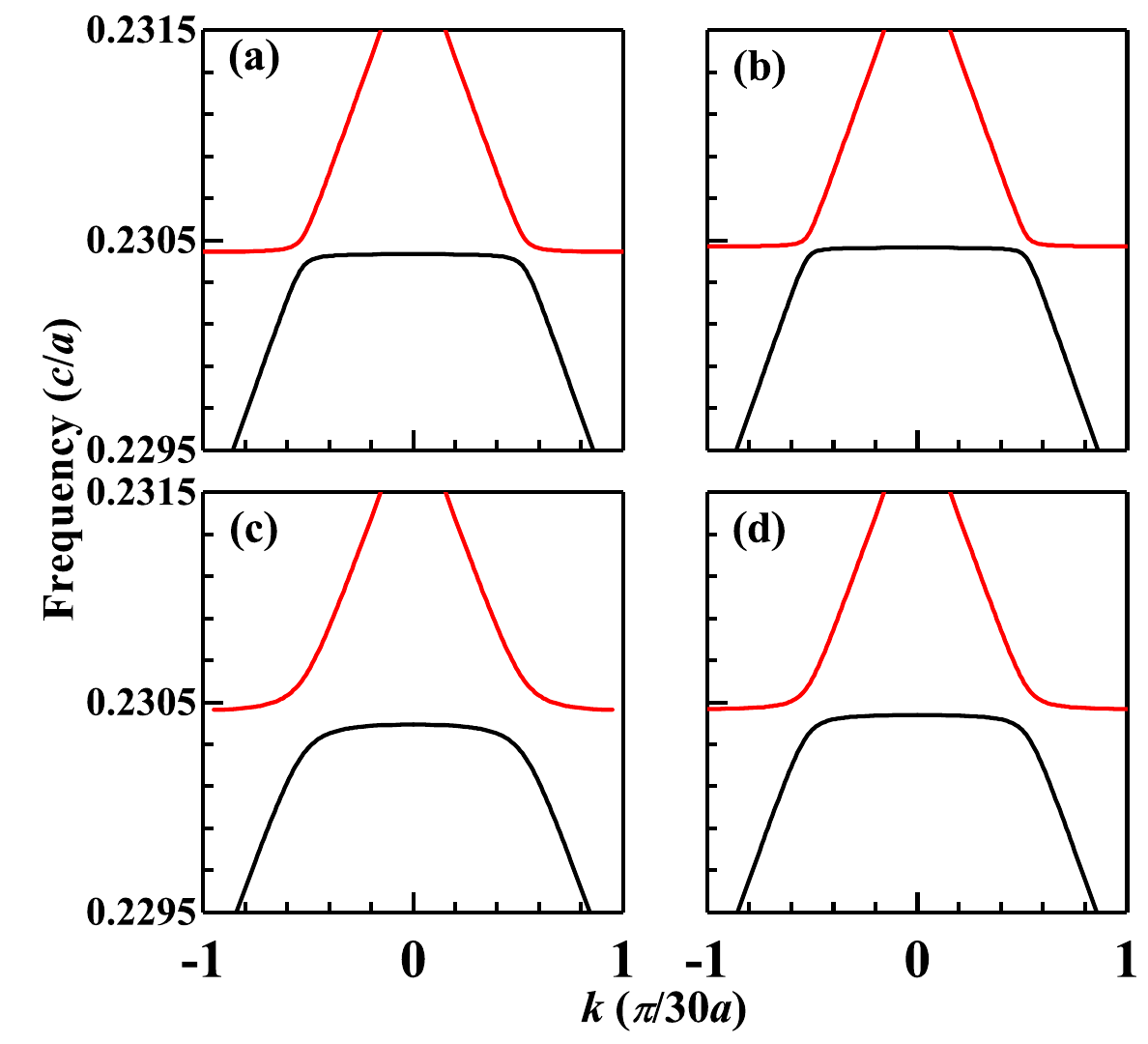}
\caption{Band structures for the super cells with different cavities, corresponding to Figs.~\ref{cavity1}(a) to \ref{cavity1}(d) in the manuscript, respectively. The periodic direction of the super cells, along the interface between the two photonic crystals, has the period of $30a$, while the total length normal to the interface is cut to be $26\sqrt{3}a$. The cavity mode can break the linear dispersion of the PTESs at its eigenfrequency. In other words, the dispersions of the PTESs with different pseudo spins are connected by the cavity levels, which is the reason why the cavity mode can flip the pseudo spin of the PTESs.}
\label{band}
\end{figure}

In order to tune the transport of the PTESs, two
types of single-mode optical cavities are considered, namely, $\bigtriangledown$-
and $\bigtriangleup$-type cavities, as shown in Fig.~\ref{cavity1}
where all cavities are achieved by deleting two bigger and one smaller air rods. There are 6 layers for Figs.~\ref{cavity1}(a)
and \ref{cavity1}(c) and 8 layers for Figs.~\ref{cavity1}(b) and
\ref{cavity1}(d) between the cavities and interface. The electric
field distributions in the insets display the cavity modes whose
eigenfrequencies are $\sim0.23045c/a$. Since the transmissivity
can decrease to zero as the incident wave frequency is around the
cavity eigenfrequency (see Fig.~\ref{cavity1}), the non-trivial
topology of the edge states is broken, which is attributed to the breaking of the $C_{6v}$ symmetry around the cavity position. The pseudospin up and down PTESs are mixed, see Fig.~\ref{band}. Note that a magnetic defect can destroy back-scattering-immune helical edge states in electronic QSHEs, but generally it cannot suppress the conductance to zero \cite{Kurilovich2017impurities,Vezvaee2018impurities}. Therefore, the transport control of the PTESs in the present system is superior to the magnetic defects in electronic QSHEs. Because the electromagnetic wave with the frequency away from the
cavity eigenfrequency cannot resonantly couple
to the cavity mode, the transmissivity is also able to approach
100\%, indicating that the breaking of the topology of the PTESs
only appears around the cavity eigenfrequency.

The transmission spectra in Fig.~\ref{cavity1} depend on
the cavity shape and distance to the PTW. To understand this, we build the PTW-cavity transport theory [see Appendix B], which gives the transmission coefficient as follow \cite{shen2009theory1, zhang2014effects, wang2016dynamics},
\begin{align}
t & = 1 - {iV_R^2/v_g \over \left(\omega-\omega_c \right)+i{V_R^2+V_L^2\over 2 v_g}}, \label{tt}
\end{align}
where $\omega$ is the frequency of the incident
wave with the group velocity $v_{g}$. $\omega_{c}$ measures the
eigenfrequency of the cavity mode whose couplings with
the rightward and leftward moving PTESs are described by the $\delta$ functions of $V_{R}\delta(x)$ and $V_{L}\delta(x)$, respectively. For the cases in Fig.~\ref{cavity1} we have $v_{g}=0.02711c$
and $V_{R}=V_{L}$. The relation of $V_{R}=V_{L}$ dates from the structure symmetry which leads to that the cavity modes hold the same weights for the pseudospin up and down states.
Considering this relation equation (\ref{tt}) gives the zero transmission
when $\varepsilon=\omega_{c}$, implying that the cavity flips the
pseudospin of the incident edge state. In order to get the coupling between the cavity modes and PTESs, we fit the transmission spectra with Eq.~(\ref{tt}) [see
the red curves in Fig.~\ref{cavity1}]. For convenience we denote $J=V_R^2/v_g\ \left(=V_L^2/v_g\right)$. The fitted values
of $(\omega_{c},\ J)$ are $\left(0.23044,\ 5.73\times10^{-6}\right)\frac{c}{a}$,
$\left(0.23047,\ 3.18\times10^{-6}\right)\frac{c}{a}$, $\left(0.23043,\ 3.65\times10^{-5}\right)\frac{c}{a}$, and $\left(0.23046,\ 1.50\times10^{-5}\right)\frac{c}{a}$ from Figs.~\ref{cavity1}(a)
to \ref{cavity1}(d), for which the fitting precisions are measured
by the $R$-square, $R^2$. Since $R^2\sim1$, it proves to be
reasonable to assume the $\delta$ coupling between the cavity
modes and PTESs. In detail, $R^2$ is a little less for Figs.~\ref{cavity1}(a) and \ref{cavity1}(c) compared with other two cases, indicating that the non-$\delta$ coupling effects appear between the PTW and cavity \cite{zhang2014effects}. For example, the red line is a little higher and lower than the circle dots on the left and right
sides of the transmission dip in Fig.~\ref{cavity1}(c). The different width and slightly different position of the dips in Figs.~\ref{cavity1}(a) and \ref{cavity1}(c) [Figs.~\ref{cavity1}(b) and \ref{cavity1}(d)] are due to the different distributions of the cavity modes toward the PTW. As the distance between the cavity and PTW increases, the overlap between the cavity modes and PTESs decreases and so does the coupling $J$ [comparing Figs.~\ref{cavity1}(a)
with \ref{cavity1}(b) or Figs.~\ref{cavity1}(c) with \ref{cavity1}(d)].
Therefore, well designed optical cavities can tune the transport of the PTESs.

\begin{figure}
\centering
\includegraphics[width=0.46\textwidth]{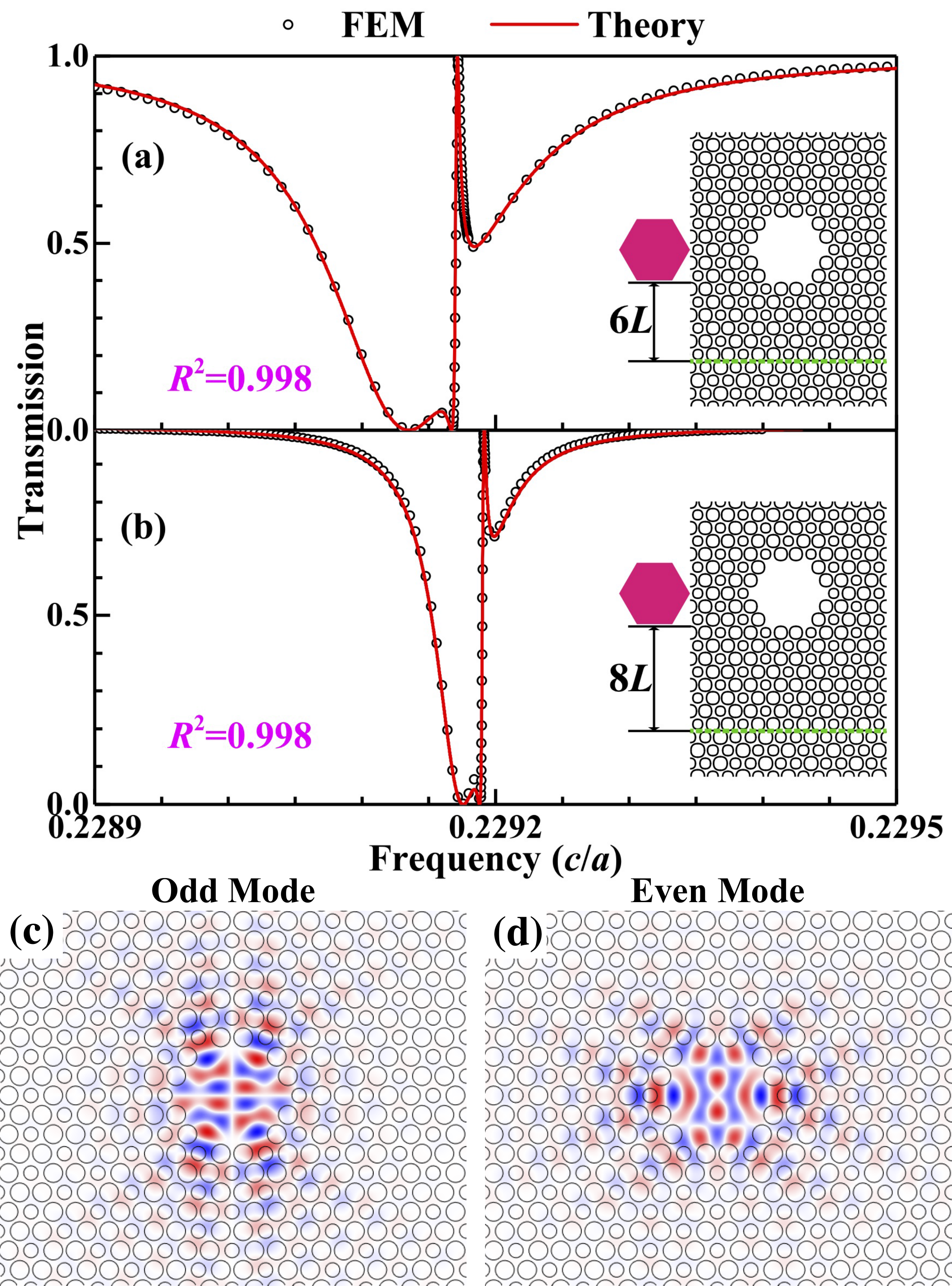}
\caption{Transmission spectra of the topological edges that are coupled with
side twofold degenerate modes cavities. Notations
of $6L$ and $8L$ show the distances between the cavities and edges,
i.e., there are 6 and 8 layers, respectively. Circle dots are calculated
from the COMSOL code, while red lines are the fitted curves whose
fitting precisions are described by the $R$-square.}
\label{cavity2}
\end{figure}

Compared with single-mode (non-degenerate) cavities in Fig.~\ref{cavity1}, the cavities with degenerate modes (i.e., degenerate cavities) can induce more diversities for the transmission spectra of the PTESs, see Figs.~\ref{cavity2}(a) and \ref{cavity2}(b) where the hexagon cavities contain twofold degenerate modes and are topological trivial. Their parities are even and odd, see their electric field distributions in Figs.~\ref{cavity2}(c) and \ref{cavity2}(d). Because the triangle cavities in Fig.~2 are achieved by deleting two bigger and one smaller air rods, they only have the mirror symmetry with respect to the vertical line through their center. As a result, their levels are commonly non-degenerate. However, the symmetry of the hexagonal cavities in Fig.~3 is described by the point group of $C_{6v}$ (the fixed point is the cavity center) which has two-dimensional representation. Therefore, the levels of the hexagonal cavities can be double degenerate.

\begin{figure}[ht]
\centering
\includegraphics[width=0.48\textwidth]{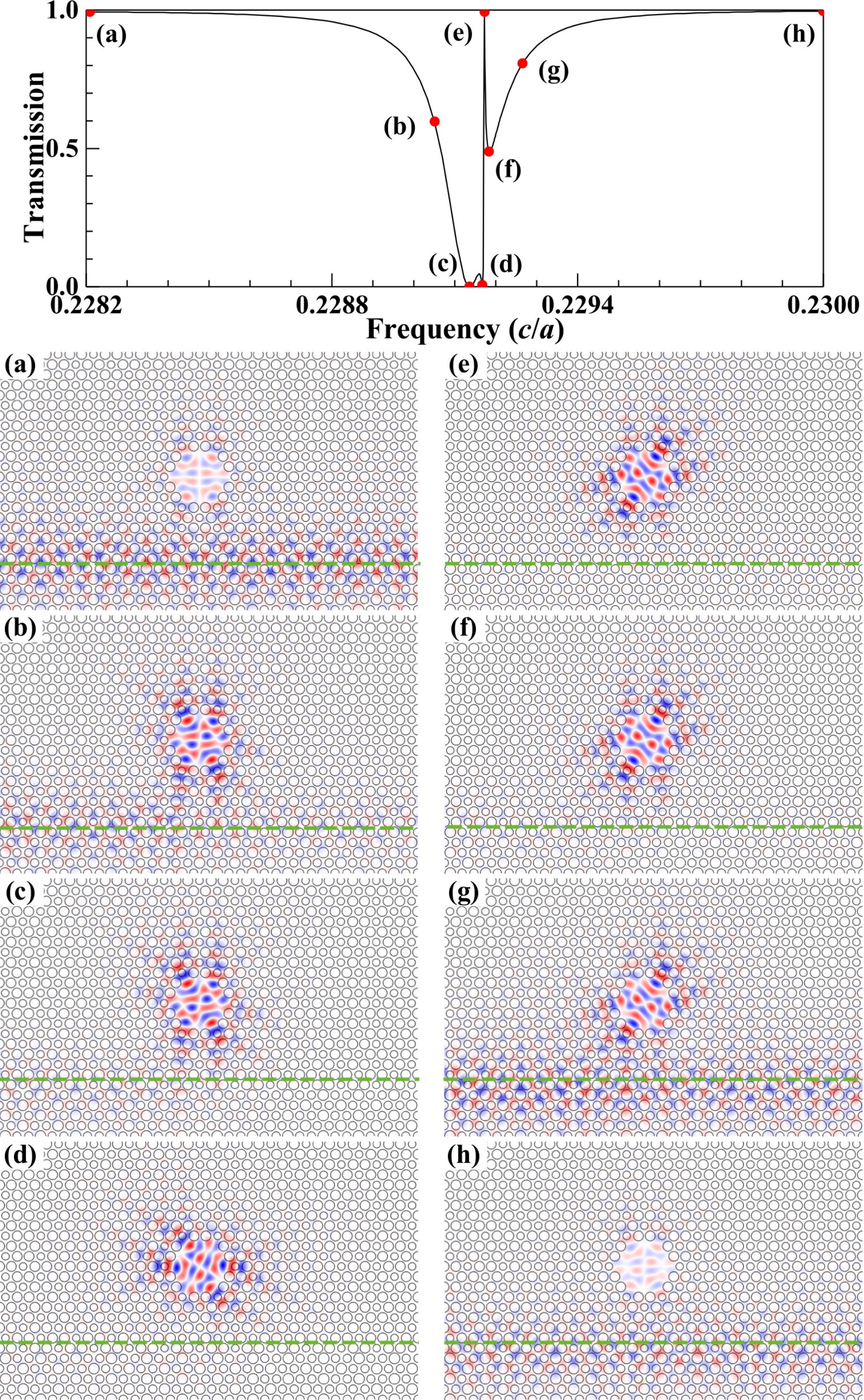}
\caption{Top panel: Same transmission curve with that in Fig.~\ref{cavity2}(a). Contour maps (a-h): Electric field distributions corresponding to the red dots of (a-h), respectively. Note that the photonic structure is the same with that used for Fig.~\ref{cavity2}(a).}
\label{6L}.
\end{figure}

\begin{figure}
\centering
\includegraphics[width=0.48\textwidth]{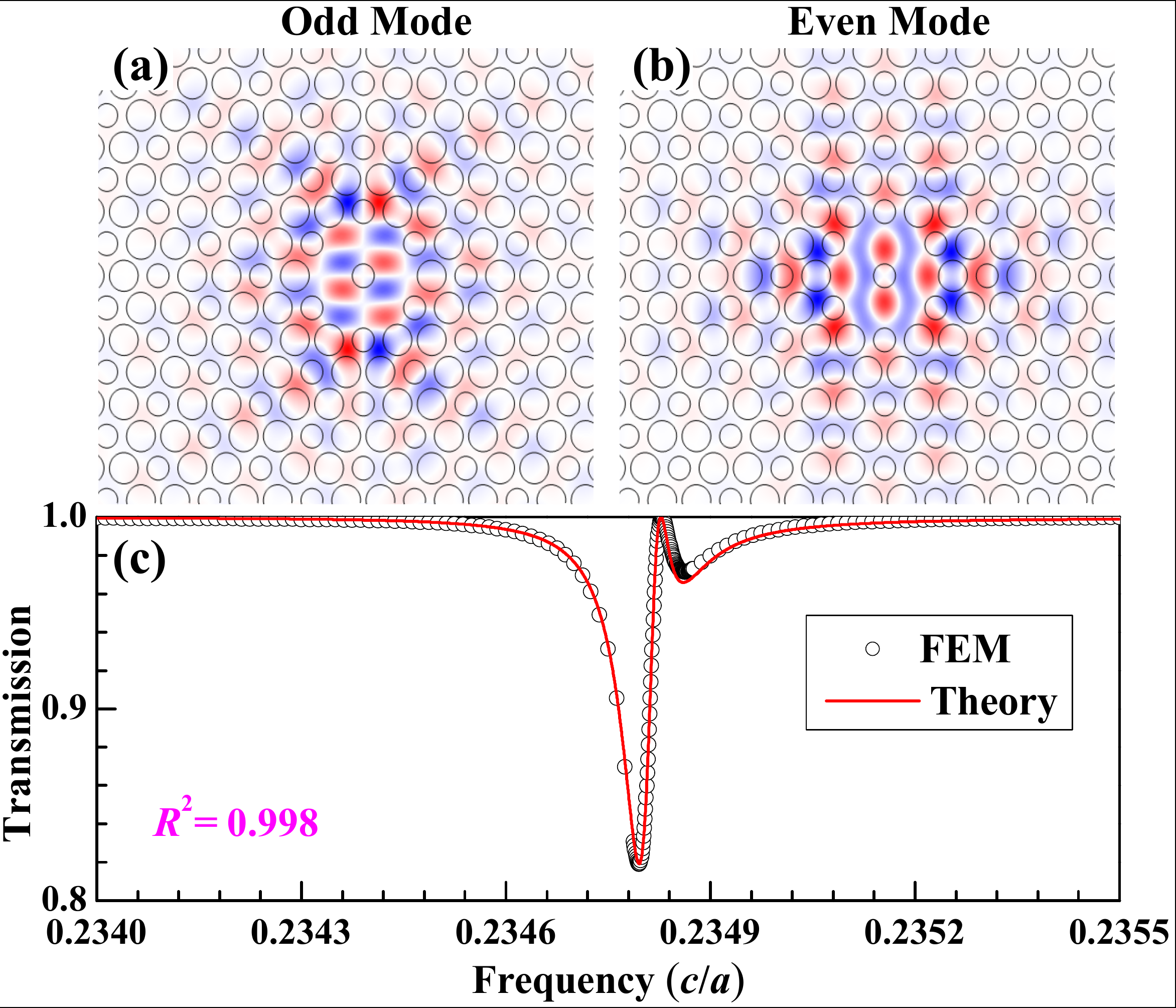}
\caption{Electric field distributions of (a) odd and (b) even modes in a different  degenerate hexagon cavity. (c) Transmission spectrum for which the distance of the cavity relative to the PTW is 9 layers. In (c), circle dots are calculated from the FEM, while the red line is the fitted one by the one-dimensional photonic transport theory for the twofold degenerate cavities built up in the above section. The fitting precision of the red line is described by the $R$-square, and the fitting parameters $(\omega_c, V_0^e, V_0^o, w_e, w_o)$ are $(0.23482{c\over a}, 4.27\times10^{-4}{c\over a}, 1.52\times10^{-4}{c\over a}, 3.02a, 18.07a)$. Note that the transmission spectrum shows a Fano line shape, indicating that it is possible to achieve many different transmission lineshapes for the PTESs by using side cavities.}
\label{fano}
\end{figure}

Since the cavities used are much larger than $a$, the couplings of their degenerate modes with the PTWs should be the non-$\delta$ type, being even (odd) for even (odd) modes \cite{zhang2014effects}. Accordingly, we develop the one-dimensional PTW-cavity transport theory to account for the effect of the degenerate cavity with non-$\delta$ coupling. It leads to more complexity with respect to that for the non-degenerate cavities with $\delta$ coupling, see Appendices B and C. We take the following coupling functions,
\begin{align}
V_e(x) =& V_0^e\left(\pi w_e^2\over4\right)^{-{1\over4}}e^{-{2x^2\over w_e^2}}, \\
V_o(x) =& V_0^o\left(\pi w_o^2\over4\right)^{-{1\over4}}{\sqrt{8}x \over w_o}e^{-{2x^2\over w_o^2}},
\end{align}
where $V_0^e$ and $w_e$ ($V_0^o$ and $w_o$) denote the coupling strength and width of the even (odd) modes, respectively. Since $V(x)$ is determined by the cavity mode, its square has been normalized to $V_0^2$ \cite{zhang2014effects}. These two non-$\delta$ functions can well describe the coupling between the cavity and PTW, referred to the red fitted curves in Fig.~\ref{cavity2} [both have $R^2= 0.998$]. Each of the degenerate modes has the same coupling strengths with the rightward or leftward moving PTESs, similar to the non-degenerate cavity. According to the theory in Appendix C, the fitted values of $(\omega_c, V_0^e, V_0^o, w_e, w_o)$ are $(0.22916{c\over a}, 7.01\times10^{-4}{c\over a}, 8.82\times10^{-4}{c\over a}, 2.50a, 0.94a)$ for Fig.~\ref{cavity2}(a) and $(0.22919{c\over a}, 3.92\times10^{-4}{c\over a}, 6.29\times10^{-4}{c\over a}, 2.78a, 1.07a)$ for Fig.~\ref{cavity2}(b). As the distance of the cavity relative to the PTW increases, the coupling strength decreases for both modes, while the widths show a very small changing, because the distribution shapes of the modes do not change much along the PTW except the strength. The wider distribution of the odd (even) mode along the vertical (horizonal) direction than that of the even (odd) one is responsible for $V_0^o > V_0^e$ ($w_e>w_o$), referred to the mode distributions in Figs.~\ref{cavity2}(c) and \ref{cavity2}(d). The transmission spectra appear as a continuous transmission dip interrupted by a single mode, resulting in a Fano line shape, see Figs.~\ref{cavity2}(a) and \ref{cavity2}(b). According to the variation of the cavity field distribution along the transmission curves [see Fig.~\ref{6L}], the dip is mainly from the odd mode while the Fano line shape from the even mode, which is due to $V_0^o>V_0^e$. In Fig.~\ref{fano}, we also give a case that a small hexagon cavity with two degenerate modes can lead to a spectrum with better Fano line shape, agreeing well with the one-dimensional PTW-cavity transport theory too.

\begin{figure}
\centering
\includegraphics[width=0.48\textwidth]{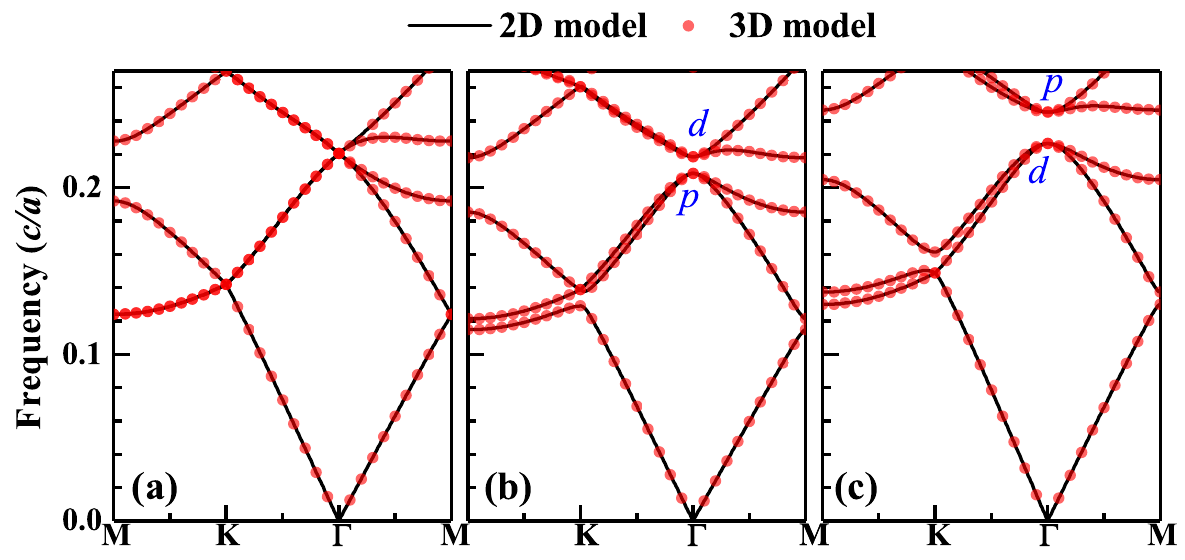}
\caption{Band structures for the lattices with rods of infinite length (black solid lines) and finite length (circular red dots). Black lines in (a-c) are the same with those lines in Figs.~\ref{structures}(d-f). For all circular red dots the length of the rods is $0.5a$. The lattice structures in (a-c) are the same with those in Figs.~\ref{structures}(a-c), respectively. The bands given by the black lines and circular red dots are consistent with each other.}
\label{bandComparison}
\end{figure}

\begin{figure}
\centering
\includegraphics[width=0.48\textwidth]{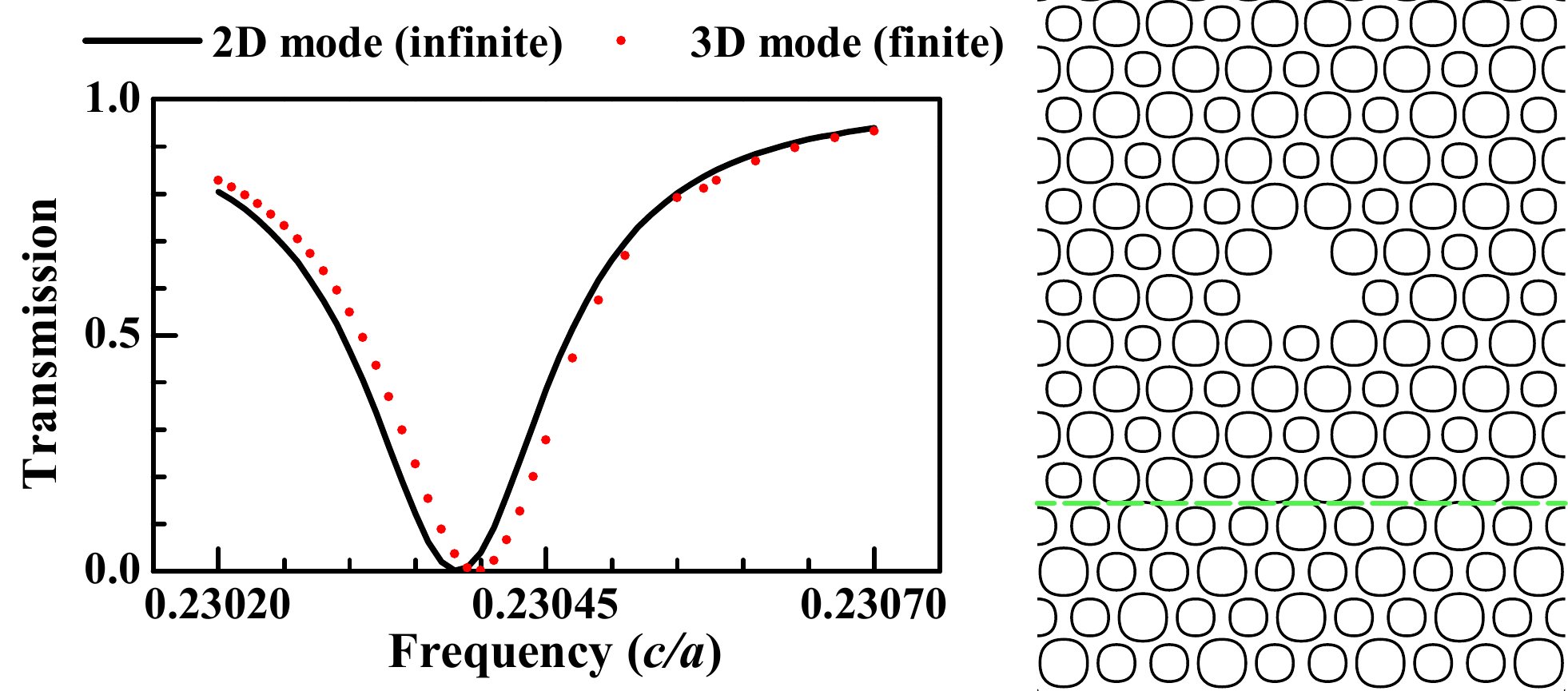}
\caption{Left panel: Transmission spectra of the topological edge coupled with side single-mode cavity for the lattices with rods of infinite length (black solid lines) and of finite length (circular red dots). For all circular red dots the length of the rods is $0.5a$. The transmission spectra given by the black line and circular red dots are consistent with each other. Right panel: Draft structure of the calculated.}
\label{TComparison}
\end{figure}

For a realistic system, since the rods of finite length may introduce longitudinally-dependent states in the topological band gap, short rods are preferred \cite{wu2015scheme, gorlach2018far, yves2017crystalline, barik2018topological}. When the length of the rods are set to $0.5a$, the conclusions of Figs.~\ref{cavity1} and \ref{cavity2} do not change, which is confirmed by Figs.~\ref{bandComparison} and \ref{TComparison}. They, respectively, show that the bands and transmission are consistent for the two cases, i.e., finite length rods  $(0.5a)$  and infinite length rods. If nonlinear optical materials are introduced into the cavities, more adjustability can be achieved for the transport of the PTESs, such as topological all-optical switches \cite{soljavcic2004enhancement, notomi2007nonlinear, nozaki2010sub, volz2012ultrafast}. One of its merits is the significant drop for signal loss when the switch is on, and another is the perfect reflection when the switch is off. Photonic crystal cavities allow a high field enhancement (beneficial for shifting the resonant frequency of the cavity) and therefore, this topological all-optical switch is possible. Similar discussion is also suitable for other topological optical devices, such as filters and logical gates.

\section{conclusion}

In conclusion, we studied the transport of the topological edge states in the crystalline-symmetry-protected photonic topological insulators. Since the photonic topological insulators are designed by air rods in conventional dielectric materials, it is practical and convenient to achieve them in experiments. The transport property of the PTESs is investigated under two kinds of defects. For the interface with different bending angles, the transmission spectra show that the edge state is robust. While for the cavity defects breaking the crystal
symmetry, the cavity modes can strongly couple with the edge states
near the resonant frequency of the cavity, resulting in
a pseudo-spin flipping and consequent reflection of the topological edge states. This phenomenon is explained by the one-dimensional
PTW-cavity transport theory that we build. The propagation of the PTESs can be easily tuned by the geometry and distance of the cavity relative to the interface and therefore, it is convenient to achieve many types of transmission line shapes, holding potential applications in integrated
optics. If a nonlinear cavity is considered, one can expect more
adjustability for the transport of the topological edge states, for
example, topological all-optical switches, filters, and logic gates.

\section*{Acknolodgement}
We thank professor Chunyin Qiu for discussion on the transmission spectral calculation. This work was supported by the National Natural Science Foundation of China (Grant Nos. 11304015, 11734003) and National Key R$\&$D Program of China (Grant Nos. 2016YFA0300600, 2017YFB0701600).

\appendix

\section{Scattering matrix method}

\begin{figure}
\centering
\includegraphics[width=0.47\textwidth]{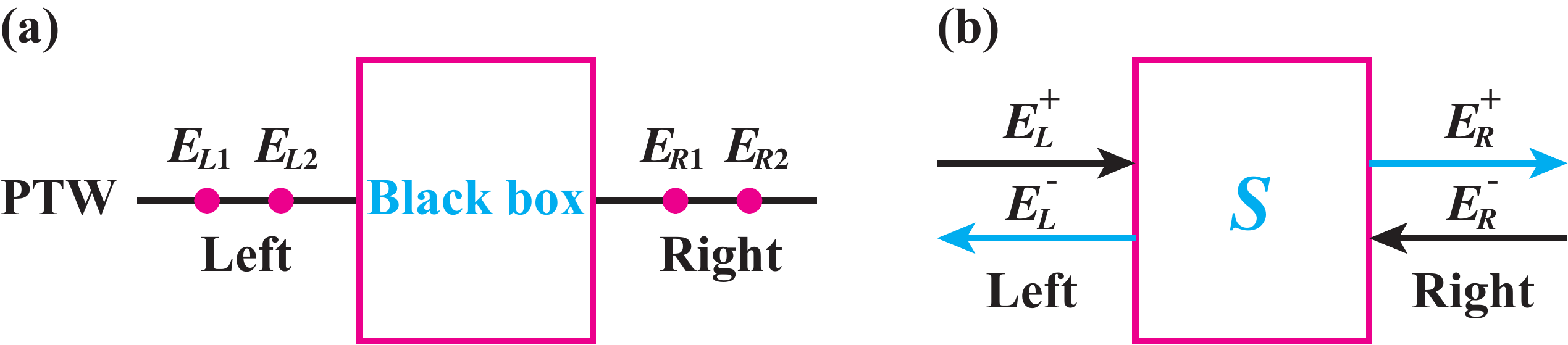}
\caption{Scattering model for one-dimensional photonic topological waveguides (PTWs). The black box in (a) is the scattering region which could be Z-type edges, cavities, or other optical defects, described by the scattering matrix $S$ in (b). The pseudospin up and down fields on the left (right) side, $E_L^+$ and $E_L^-$ ($E_R^+$ and $E_R^-$), can be abstracted from those fields on the left (right) two spatially equivalent points, $E_{L1}$ and $E_{L2}$ ($E_{R1}$ and $E_{R2}$).}
\label{smt}
\end{figure}

Throughout the work, we use the scattering matrix method to calculate the transport of topological edge states coupled with optical defects. This method has been used to calculate the topological valley transport of sound in sonic crystals \cite{lu2016valleytransport}. The defects in the scattering matrix method are taken as the black box [see Fig.~\ref{smt}(a)] which connects the left and right channels. The fields on two spatially equivalent points in each channel can provide the rightward and leftward moving (i.e., pseudospin up and down) field components. In Fig.~\ref{smt}, the rightward and leftward moving fields of $E_L^+$ and $E_L^-$ ($E_R^+$ and $E_R^-$) can be found by $E_{L1}$ and $E_{L2}$ ($E_{R1}$ and $E_{R2}$), namely,
 \begin{align}
E_{L1} &= E_L^++E_L^-,\\
E_{L2} &= E_L^+e^{ikd}+E_L^-e^{-ikd},\\
E_{R1} &= E_R^++E_R^-,\\
E_{R2} &= E_R^+e^{ikd}+E_R^-e^{-ikd}.
\end{align}
where $k$ is the Bloch wave vector and $d$ is the integral multiple of the lattice period along the waveguide. On the other hand, the scattering waves of $E_L^-$ and $E_R^+$ can be expressed by the incident waves of $E_L^+$ and $E_R^-$ with ths scattering matrix of $S$,
\begin{equation}
\begin{bmatrix}
E_L^-\\ E_R^+
\end{bmatrix}
=S
\begin{bmatrix}
E_L^+\\ E_R^-
\end{bmatrix}
\end{equation}
where
\begin{equation}
S=\begin{bmatrix}
r & t \\
t & -\frac{r^*t}{t^*}\\
\end{bmatrix}.
\label{Seq}
\end{equation}
Here, $r$ and $t$ are the reflection and transmission coefficients of the pseudo-spin states, respectively. Note that the matrix $S$ expression in Eq.~\eqref{Seq} requires the energy conservation ($S^\dag S = I$) and time reversal symmetry ($S^*S=I$), both of which are satisfied by our photonic systems.

\section{Transport theory for Single mode cavity}

In this section, we give the derivation of the Eq.~(3) in the manuscript for understanding the influence of the side cavity on the transmission of the photonic topological waveguides (PTWs). The coupled architecture of one-dimensional PTW with the side cavity can be described by the following Hamiltonian \cite{shen2009theory1},
\begin{align}
H=H_W+H_{C}+H_{I}
\label{eq:ham}
\end{align}
where $H_W$ and $H_{C}$ are the Hamiltonians of the waveguide and cavity, respectively, and $H_{I}$ represents their coupling. They can be written as,
\begin{align}
H_W =& \int dx\hat L^\dag (x)\left(\omega_c-v_gk_c+iv_g \frac {\partial}{\partial x}\right) \hat L(x) \nonumber \\
& + \int dx {\hat R}^\dag (x) \left(\omega_c-v_gk_c-iv_g \frac {\partial}{\partial x}\right) \hat R(x),\\
H_{C}=&\ \omega_c \hat c^\dag \hat c,\\
H_{I}=& \int dx V_R\delta(x)\left[c^\dag\hat R(x)+{\hat R}^\dag (x)\hat c \right] \nonumber \\
& + \int dx V_L\delta(x)\left[c^\dag\hat L(x)+{\hat L}^\dag (x)\hat c \right],
\label{eq:ham_i}
\end{align}
where ${\hat R}^\dag(x)$ and ${\hat L}^\dag(x)$ $\left[{\hat R}(x)\right.$ and $\left.{\hat L}(x)\right]$ are the photon field creation [annihilation] operators of the rightward- and leftward-moving waveguide modes, corresponding to the pseudo-spin up and down photonic topological edge states (PTESs), respectively. $\hat c^\dag$ is the creation operator of the cavity mode with eigenfrequency of $\omega_c$. Near $\omega_c$ the dispersions of the rightward and leftward moving PTESs are linear with respect to the wave vector $k$, i.e., $\omega_\pm=\omega_c-v_g k_c \pm v_g k$, where $v_g$ is the group velocity and $k_c$ is determined by $\omega_+|_{k=k_c}=\omega_c$. Since the distribution of the considered cavity modes along the PTW is smaller than the wavelength of the incident light, we assume a $\delta$-type coupling between cavity and PTW. The $\delta$ functions of $V_R\delta(x)$ and $V_L\delta(x)$ describe the coupling of the cavity mode with the pseudo-spin up and down PTESs, respectively. Generally, we have $V_{R}=V_{L}$ due to the structure symmetry, that is, the cavity modes hold the same weights for the pseudospin up and down states..

The system eigenstate is in the following form,
\begin{align}
|\Phi\rangle &= \int dx{\cal R}(x){\hat R}^\dag(x)|\varnothing\rangle
+ \int dx{\cal L}(x){\hat L}^\dag(x)|\varnothing\rangle
+{\cal C}c^\dag|\varnothing\rangle
\label{eq:wf}
\end{align}
where $|\varnothing\rangle$ represents the vacuum state, with zero photon in the cavity and waveguide. $\cal C$ is the excitation amplitude of the optical cavity. ${\cal R}(x)$ and ${\cal L}(x)$ are the photon wave functions of the pseudo-spin up states and down states, respectively. Substituting Eqs.\ \eqref{eq:ham} and \eqref{eq:wf} into the steady state Schr\"odinger equation,
\begin{align}
H|\Phi\rangle=\omega |\Phi\rangle
\label{eq:seq}
\end{align}
we can get the coupled equations for ${\cal R}(x)$, ${\cal L}(x)$ and $C$ as follows:
\begin{subequations}
\begin{align}
-iv_g\frac\partial{\partial x}{\cal R}(x)+V_R\delta(x){\cal C}&=(\omega-\omega_c+v_gk_c){\cal R}(x),\\
 iv_g\frac\partial{\partial x}{\cal L}(x)+V_L\delta(x){\cal C}&=(\omega-\omega_c+v_gk_c){\cal L}(x),\\
V_R{\cal R}(0)+V_L{\cal L}(0)&=(\omega-\omega_c){\cal C}.
\label{eq:rlc}
\end{align}
\end{subequations}
Adopting the following wave functions for the pseudo-spin up and down states,
\begin{align}
{\cal R}(x)=e^{ikx}\theta(-x)+te^{ikx}\theta(x),\quad {\cal L}(x)= re^{-ikx}\theta(-x),
\label{eq:seq8}
\end{align}
one can find the transmission coefficient, $t$, namely,
\begin{align}
t  = 1 - {iV_R^2/v_g \over \left(\omega-\omega_c \right)+i{V_R^2+V_L^2\over 2 v_g}}.
\label{eq:tt}
\end{align}
This is the equation (3) in the manuscript.

\section{Transport theory for twofold degenerate cavities}

In the present section, we show the photonic transport theory for the PTW coupled with a twofold degenerate cavity. This photonic topological system can be described by the following Hamiltonian,
\begin{align}
H=H_W+H_{C}+H_{Io}+H_{Ie},
\label{eq:ham2}
\end{align}
where
\begin{align}
H_W =& \int dx\hat L^\dag (x)\left(\omega_c-v_gk_c+iv_g \frac {\partial}{\partial x}\right) \hat L(x)\nonumber \\
& + \int dx {\hat R}^\dag (x) \left(\omega_c-v_gk_c-iv_g \frac {\partial}{\partial x}\right) \hat R(x),\\
H_{C} =&\omega_c (\hat c^\dag_{o} \hat c_{o}+\hat c^\dag_{e} \hat c_{e}),\\
H_{Io} =& \int dx V_{Ro}(x)\left[c^\dag_{o}\hat R(x)+{\hat R}^\dag (x)\hat c_{o} \right] \nonumber \\
& + \int dx V_{Lo}(x)\left[c^\dag_{o}\hat L(x)+{\hat L}^\dag (x)\hat c_{o} \right],\\
H_{Ie} =& \int dx V_{Re}(x)\left[c^\dag_{e}\hat R(x)+{\hat R}^\dag (x)\hat c_{e} \right]\nonumber \\
& + \int dx V_{Le}(x)\left[c^\dag_{e}\hat L(x)+{\hat L}^\dag (x)\hat c_{e} \right].
\end{align}
$\hat c^\dag_{o}$ and $\hat c^\dag_{e}$ ($\hat c_{o}$ and $\hat c_{e}$) are the creation (annihilation) operators of the twofold degenerate modes with odd and even parities, respectively, whose eigenfrequencies both are $\omega_c$. Note that the considered twofold degenerate cavities in the present work are much larger than $a$. Therefore, we use the non-$\delta$ functions of $V_{Ro}(x)$ and $V_{Lo}(x)$ [$V_{Re}(x)$ and $V_{Le}(x)$] to describe the coupling between the PTW and odd [even] cavity mode. Since the cavity modes determine the parities of the coupling functions, $V_{Ro}(x)$ and $V_{Lo}(x)$ should be odd functions, while $V_{Re}(x)$ and $V_{Le}(x)$ are even ones \cite{zhang2014effects}. We also have $V_{Ro}(x)=V_{Lo}(x)$ and $V_{Re}(x)=V_{Le}(x)$ due to the structure symmetry. The meanings of all other symbols are the same with those in Eqs.~\eqref{eq:ham}-\eqref{eq:ham_i}. Similar to Eq.~\eqref{eq:wf}, the wave function in the present system is
\begin{align}
|\Phi\rangle = & \int dx{\cal R}(x){\hat R}^\dag(x)|\varnothing\rangle
+ \int dx{\cal L}(x){\hat L}^\dag(x)|\varnothing\rangle\nonumber \\
& +{\cal C}_{o}c^\dag_{o}|\varnothing\rangle+{\cal C}_{e}c^\dag_{e}|\varnothing\rangle
\label{eq:wf2}
\end{align}
where ${\cal C}_o$ and ${\cal C}_e$ are the excitation amplitudes of the odd and even cavity modes, respectively. Substituting Eqs.~\eqref{eq:wf2} and \eqref{eq:ham2} into Eq.~\eqref{eq:seq}, we get the equation set,
\begin{subequations}
\begin{align}
-iv_g\frac\partial{\partial x}{\cal R}(x)+V_{Re}(x){\cal C}_{e}+V_{Ro}(x){\cal C}_{o}&=(\omega-\omega_c+v_gk_c){\cal R}(x),\\
 iv_g\frac\partial{\partial x}{\cal L}(x)+V_{Lo}(x){\cal C}_{o}+V_{Le}(x){\cal C}_{e}&=(\omega-\omega_c+v_gk_c){\cal L}(x),\\
 \int dx[{\cal R}(x)V_{Ro}(x)+V_{Lo}(x){\cal L}(x)]&=(\omega-\omega_c){\cal C}_{o},\\
\int dx[{\cal R}(x)V_{Re}(x)+V_{Le}(x){\cal L}(x)]&=(\omega-\omega_c){\cal C}_{e}.
\end{align}
\label{eq:dyn2}
\end{subequations}
for ${\cal R}(x)$, ${\cal L}(x)$, $C_o$, and $C_e$.

In order to find the transmission coefficient, one can make the following transform
\begin{align}
{\cal R}(x)=\xi_R (x)e^{ikx},\quad
{\cal L}(x)= \xi_L (x)e^{-ikx}
\label{eq:rlxi}
\end{align}
for ${\cal R}(x)$ and ${\cal L}(x)$. The corresponding boundary conditions for $\xi_R (x)$ and $\xi_L (x)$ are:
\begin{align}
\xi_R(-\infty)=1, \ \xi_R(\infty)=t, \ \xi_L(-\infty)=r, \ \xi_L(\infty)=0.
\label{eq:bds}
\end{align}
where $t$ and $r$ are the transmission and reflection coefficients, respectively. Considering $V_{Ro}(x)=V_{Lo}(x)$ and $V_{Re}(x)=V_{Le}(x)$, we denote them as $V_o(x)$ and $V_e(x)$. Using Eqs.~\eqref{eq:rlxi} and \eqref{eq:bds}, the equation set \eqref{eq:dyn2} changes into
\begin{subequations}
\begin{align}
 (\omega-\omega_c-\Delta_k^{oo}){\cal C}_o-\Delta_k^{oe}{\cal C}_e-V_k^o r&=V_{-k}^o,\\
(\omega-\omega_c-\Delta_k^{ee}){\cal C}_e-\Delta_k^{eo}{\cal C}_o-V_k^e r&=V_{-k}^e,\\
 iv_g^{-1}\left[V^o_{-k}{\cal C}_o+V^e_{-k}{\cal C}_e\right]+r&=0,\\
 iv_g^{-1}\left[V^e_k{\cal C}_e+V^o_k{\cal C}_o\right]+ t&=1.
\end{align}
 \label{eq:lineq}
\end{subequations}
Here,
\begin{subequations}
\begin{align}
V^m_k&=\int_{-\infty}^\infty dx' V_m(x' )e^{-ikx' },\\
\Delta_k^{mn} &= {2\over v_g} \int_0^\infty dx Q_{mn}(x)\sin(kx), \\
 Q_{mn}(x) &= \int_{-\infty}^{\infty} dx'[V_m(x')V_n(x'-x)],
\end{align}
\label{eq:coes}
\end{subequations}
where $m$ and $n$ are in $\{o,\ e\}$. The equation \eqref{eq:lineq} is the linear equation set of ${\cal C}_o$, ${\cal C}_e$, $r$, and $t$, where all parameters are given by Eq.~\eqref{eq:coes}. Therefore, one can find the transmission coefficient from Eq.~\eqref{eq:lineq}, once the coupling functions of $V^o(x)$ and $V^e(x)$ are provided. In the present work, they are assumed as \cite{zhang2014effects}
\begin{subequations}
\begin{align}
V_e(x) &= V_0^e\times\left({1\over4}\pi w_e^2\right)^{-1/4}e^{-2x^2/w_e^2},\label{Ve}\\
V_o(x) &= V_0^o\times\left({1\over4}\pi w_o^2\right)^{-1/4}\sqrt{8}{x \over w_o}e^{-2x^2/w_o^2},\label{Vo}
\end{align}
\end{subequations}
where $V_0^e$ and $w_e$ ($V_0^o$ and $w_o$) denote the coupling strength and width of the even (odd) modes, respectively. These parameters are determined by fitting the transmission calculated from the FEM with $|t|^2$ obtained from Eq.~\eqref{eq:lineq}. The expression of $|t|^2$ is not shown here, for its length is too long.

\bibliographystyle{apsrev4-2}

%

\end{document}